\RequirePackage{ifpdf}
\documentclass[12pt]{article}

\usepackage{epsfig,graphicx,bm,amsmath,amsfonts}
\usepackage{epstopdf}
\usepackage{amssymb,xcolor,float}
\usepackage{amsthm}
\usepackage{hyperref}
\usepackage{cite}
\input epsf.sty
\usepackage{cite}
\usepackage{comment}
\usepackage{epstopdf}
\usepackage{float}
\usepackage{caption}
\usepackage{subcaption}
\usepackage{enumitem}
\usepackage{bbm}
\usepackage{bm}
\usepackage{longtable}
\usepackage{parskip}
\usepackage{enumitem}
\usepackage{graphicx,epsfig}
\usepackage{ae,aecompl}
\usepackage{dcolumn}
\usepackage{latexsym}
\usepackage{psfrag}
\usepackage{ytableau}
\usepackage{tabularx}
\usepackage{float}
\usepackage{wrapfig}

\usepackage{xcolor}
\usepackage[english]{babel}
\usepackage[T1]{fontenc}
\usepackage[utf8]{inputenc}
\usepackage{authblk}
\usepackage{mathtools}
\usepackage{slashed}
\usepackage{mattens,amssymb}
\usepackage{amsmath,amssymb}
\usepackage{amsfonts}
\usepackage{graphicx,color}
\usepackage{cite}
\usepackage{hyperref}
\hypersetup{
	colorlinks=true,
	linkcolor=blue,
	filecolor=magenta,      
	urlcolor=cyan,
}
\usepackage[capitalize]{cleveref}
\numberwithin{equation}{section} 

\definecolor{refcol}{rgb}{0.9,0.1,0.1}
\hypersetup{colorlinks=true,linkcolor=blue,citecolor=refcol,urlcolor=cyan,linktocpage}

\usepackage{wrapfig}

\newcommand{\Tr}{\text{Tr}}

\newcommand{\ben}{\begin{eqnarray}\displaystyle}
\newcommand{\een}{\end{eqnarray}}

\newcommand{\be}{\begin{equation}}
\newcommand{\ee}{\end{equation}}

\newcommand{\bc}{\begin{center}}
\newcommand{\ec}{\end{center}}

\newcommand{\eesp}{\end{split}}
\newcommand{\bsp}{\begin{split}}

\newcommand{\Rmnum}[1]{\expandafter\@slowromancap\romannumeral #1@}

\renewcommand{\d}{\delta}	

\renewcommand{\o}{\omega}	
\newcommand{\q}{\theta}	
	
\renewcommand{\r}{\rho}		

\newcommand{\x}{\xi}								

\newcommand{\Z}{\mathcal{Z}}

\newcommand{\cA}{\mathcal{A}}
\newcommand{\cB}{\mathcal{B}}

\newcommand{\cH}{\mathcal{H}}

\newcommand{\cO}{\mathcal{O}}

\newcommand{\cZ}{\mathcal{Z}}

\newcommand{\ra}{\rightarrow}

\newcommand{\lB}{\left [}
\newcommand{\rB}{\right ]}
\newcommand{\lb}{\left (}
\newcommand{\rb}{\right )}

\newcommand{\tand}{\text{and}}

\newcommand{\bensp}{\begin{eqnarray}\begin{split}}
\newcommand{\eensp}{\end{eqnarray}\end{split}}

\newcommand{\bnm}{\begin{matrix}}
\newcommand{\enm}{\end{matrix}}

\def\XXint#1#2#3{{\setbox0=\hbox{$#1{#2#3}{\int}$ }
\vcenter{\hbox{$#2#3$ }}\kern-.6\wd0}}

\newcommand{\dow}{\partial}

\newcommand{\del}{\partial}

\begin{document}

\parindent=12pt

\begin{center}

{\Large \bf Higher Spin Gravity in $AdS_3$ and Folds on Fermi Surface}
\end{center}

\baselineskip=18pt

\bigskip

\centerline{Suvankar Dutta$^{a}$, Debangshu Mukherjee$^{b}$ and Sanhita Parihar$^{a}$}

\bigskip

\centerline{\large \it $^{a}$Indian Institute of Science Education and Research
Bhopal}
\centerline{\large \it Bhopal bypass, Bhopal 462066, India}
\centerline{\large \it $^{b}$Department of Physics, Indian Institute of Technology Kanpur}
\centerline{\large \it Kalyanpur, Kanpur 208016, India}

\bigskip

\centerline{E-mail: suvankar@iiserb.ac.in, debangshu@iitk.ac.in, sanhita18@iiserb.ac.in}

\vskip .6cm
\medskip

\vspace*{4.0ex}

\centerline{\bf Abstract} \bigskip

\noindent 
In this paper, we introduce new sets of boundary conditions for higher spin gravity in $AdS_3$ where the boundary dynamics of spin two and other higher spin fields are governed by the interacting collective field theory Hamiltonian of Avan and Jevicki. We show that the time evolution of spin two and higher spin fields can be captured by the classical dynamics of folded fermi surfaces in the similar spirit of Lin, Lunin and Maldacena. We also construct infinite sequences of conserved charges showing the integrable structure of higher spin gravity (for spin 3) under the boundary conditions we considered. Further, we observe that there are two possible sequences of conserved charges depending on whether the underlying boundary fermions are non-relativistic or relativistic.

\vfill\eject

\tableofcontents


\section{Introduction}
\label{sec:intro}

The dynamics of classical gravity in three spacetime dimensions has no bulk degrees of freedom, completely fixed by the asymptotic boundary conditions. Using the Hamiltonian formalism for gravity in three-dimensional $AdS$ ($AdS_3$) spacetime, Brown and Henneaux showed that for a specific choice of boundary conditions where the lapse and the shift functions are held constant at the asymptotic boundary, the asymptotic symmetry group is generated by two copies of Virasoro algebras \cite{Brown:1986nw}. Such boundary conditions are known as \emph{Brown-Henneaux} boundary conditions and are considered to be the standard boundary conditions. In the language of Chern-Simons theory Brown-Henneaux boundary conditions correspond to fixing the temporal component of the gauge field (known as chemical potential) to a constant value at the asymptotic boundary\footnote{The condition on constant chemical potential was relaxed in \cite{Henneaux:2013dra,Bunster:2014mua}}. However, gravity in $AdS_3$ admits a new family of non-standard boundary conditions leveled by a non-negative integer $k$ \cite{Perez:2016vqo}\footnote{Other consistent boundary conditions are also interesting and discussed in the literature. For example \cite{Troessaert:2013fma, Avery:2013dja, Afshar:2016wfy, Sheikh-Jabbari:2016npa, Afshar:2016kjj, Grumiller:2016kcp, Ojeda:2019xih, Ammon:2017vwt, Grumiller:2019fmp, Grumiller:2019tyl}. The most general asymptotically $AdS_3$ boundary conditions for Einstein's gravity were discussed in \cite{Grumiller:2016pqb}. Also, for $\mathcal{N}=(1,1)$ extended higher spin $AdS_{3}$ supergravity the boundary conditions has been explored in \cite{Ozer:2019nkv}.}. In such cases the chemical potentials are no longer kept fixed at the asymptotic boundary, rather they are allowed to explicitly depend on the fields (asymptotic values of the angular components of the gauge fields). The classical dynamics for these boundary conditions are governed by the $k$-th element of the KdV hierarchy. The standard Brown-Henneaux boundary conditions correspond to the special case $k=0$. For $k=1$, Einstein's equations are similar to two independent copies of KdV equations and the asymptotic symmetry algebra is infinite dimensional, without any central charge. The field theoretic realisation of these infinite dimensional symmetry algebras has been discussed in \cite{Gonzalez:2018jgp}. Another class of boundary conditions, known as \emph{soft hairy} boundary conditions were considered in \cite{Afshar:2016wfy,Afshar:2016kjj} for fixed chemical potentials at boundary. The soft hairy boundary conditions were further generalised in \cite{Ojeda:2019xih,Grumiller:2019tyl} by choosing the chemical potentials as appropriate local functions of the fields. It was shown that for this class of boundary conditions the classical dynamics of gravity is governed by the Gardner hierarchy of nonlinear partial differential equations, known as \emph{mixed KdV-mKdV} hierarchy. The story of pure gravity, discussed so far, can also be generalised to higher spin gravity (and also supergravity) in $AdS_3$ \cite{Henneaux:2010xg, Campoleoni:2010zq, Campoleoni:2011hg, Perez:2012cf, Henneaux:2013dra, Bunster:2014mua, Grumiller:2016kcp, Gutperle:2011kf, Gutperle:2014aja, Beccaria:2015iwa,Ozer:2019nkv}. In \cite{Ojeda:2020bgz} a new set boundary conditions were introduced for gravity coupled with spin three field where the boundary dynamics is governed by the members of the modified Boussinesq integrable hierarchy. See also \cite{Compere:2013gja, Gutperle:2014aja, Beccaria:2015iwa}. It is known that the KdV equation (also mKdV and Boussinesq equations) describes an integrable dynamical system due to the presence of an infinite number of conserved quantities (constants of motion). The above results thus show a connection between classical higher spin gravity in $AdS_3$ and certain special classes of integrable systems (KdV/mKdV/Boussinesq) in $1+1$ dimensions.

Another example of simple and interesting integrable system is the one dimensional matrix model of $N\times N$ unitary matrices with arbitrary potential. In the large $N$ limit, the one dimensional matrix model can be described in terms of a classical cubic collective field theory in arbitrary background potential \cite{sakita, jevicki}. The large $N$ degrees of freedom of the matrix model are captured by a real collective bosonic field, called the eigenvalue density. A generalisation of collective field theory including the interaction between the original collective fields and an infinite set of supplementary fields was proposed in \cite{Avan:1992gm}. One interesting feature of the collective field theory is that it admits a free fermi phase space description in terms of two dimensional droplets due to bosonisation \cite{polchinski, Dhar:1992rs, Dhar:1993jc, Dhar:1992si, Dhar:2005qh, Dhar:1992hr, Chattopadhyay, Chattopadhyay:2020rle}. A similar phase space description also exists for interacting collective field theory. It was shown in \cite{Das:1995gd, Das:1995jw} that the dynamics of interacting collective field theory can be described in terms of folds on fermi surfaces. It turns out that one can construct an infinite sequence of classical conserved charges (i.e. the Poisson brackets of these charges vanish) in cubic collective field theory and hence the theory admits an integrable structure \cite{Avan:1991kq, Avan:1992hv}\footnote{The classical integrability was generalised to quantum integrability in \cite{Avan:1991ik}.}. Therefore it is natural to ask whether the free as well as the interacting collective field theories have any dual gravity descriptions.

In this paper we introduce new sets of boundary conditions (similar to generalised soft hairy boundary conditions) for higher spin gravity in $AdS_3$ where the boundary dynamics of spin two and other higher spin fields is governed by the interacting collective field theory Hamiltonian \cite{Avan:1992gm}. We show that the time evolution of spin two and higher spin fields can be captured by the classical dynamics of folded fermi surfaces in the similar spirit of Lin, Lunin and Maldacena (LLM) \cite{Lin:2004nb}. This allows us to provide a free fermionic description of higher spin gravity in $AdS_3$. For spin 3 excitations we also construct infinite sequences of conserved charges showing the integrable structure of higher spin gravity under the boundary conditions we consider. In particular, denoting the single free fermion Hamiltonian density by $\mathfrak{h}(p,\theta)$, we find that the infinite sequences of conserved charges are given by different integral powers of $\mathfrak{h}(p,\theta)$ integrated over the phase space in presence of folds. Further, we observe that there are two possible sequences depending on whether the underlying fermions are non-relativistic or relativistic. We also show that the entropy of higher spin black holes connected to BTZ black holes is proportional to the total area of the droplets with folds.

The organisation of our paper is as follows. In sec. \ref{sec:cft} we review matrix quantum mechanics and the corresponding bosonic collective field theory. We also provide a phase space description of the theory and discuss the notion of folds. In sec. \ref{sec:hsgravity} we discuss about $AdS_3$ gravity coupled to higher spin fields and their corresponding Chern-Simons description. Asymptotic symmetries and the construction of conserved charges have also been summarised in this section. Droplet description of $AdS_3$ gravity coupled to higher spin is discussed in sec. \ref{sec:CFTgravity}. We construct two different sets of asymptotic conserved charges in higher spin gravity depending on whether the underlying fermions are non-relativistic or relativistic. We finally end with some discussions and future directions in sec. \ref{sec:discussion}.

\section{The collective field theory and free fermi droplets}\label{sec:cft}

In this section, we briefly review the basic ideas of matrix quantum mechanics and their connections with collective field theories. We also provide a phase space description of the theory and discuss the notion of folds.

We start with the partition function of a unitary matrix model in $(0+1)$ dimension
\ben\label{eq:mmqmech}
\Z_t = \int [DU] \exp\lB \int dt \lb \Tr\dot U^2 
+ W(U)\rb \rB
\een
where $U$ is a $N\times N$ unitary matrix, the trace is taken over fundamental representation, $W(U)$ is a gauge invariant function of $U$ and $D[U]$ is the Haar measure over $U(N)$ group manifold. The matrix model (\ref{eq:mmqmech}) admits two equivalent descriptions - one in terms of free fermions \cite{BIPZ} and the other in terms of collective bosonic field \cite{sakita, jevicki}. These two descriptions are related by bosonisation.

In a series of papers Jevicki and Sakita \cite{sakita, jevicki} showed that the matrix model (\ref{eq:mmqmech}) can be described in terms of a real collective bosonic field (the eigenvalue density)
\be
\rho(t,\theta) = \frac1N \sum_{i=1}^N \delta(\theta-\theta_i) 
\ee
where $\theta_i$s are eigenvalues of $U$ and its conjugate momentum $\pi(t,\q)$. The corresponding bosonic Hamiltonian is given by,
\be\label{eq:HamB}
H_B = \int d\q \lb \frac12 \frac{\dow \pi(t,\q)}{\dow\q} \r(t,\q)
\frac{\dow \pi(t,\q)}{\dow\q} + \frac{\pi^2 \r^3(t,\q)}{6} + W(\q)
\r(t,\q) \rb
\ee
where $W(\theta)$ is defined as 
\begin{equation}
    \int d\theta W(\theta) \rho(t,\theta)= W(U(t)).
\end{equation}
The dynamics of the collective field and its conjugate momentum is governed by Hamilton's equations,
\ben
\begin{split}\label{eq:colleom}
\dow_t \r(t,\q)+\dow_\q \lb \r(t,\q) v(t,\q)\rb & = 0\\
\dow_t v(t,\q) + \frac12 \dow_\q v(t,\q)^2 + \frac{\pi^2}2 \dow_\q
\r(t,\q)^2 & = -W'(\q),\quad \text{where} \quad v(t,\q) = \dow_\q \pi(t,\q).
\end{split}
\een
The above set of equations are coupled, non-linear and partial differential equations. It is possible to decouple these two equations by introducing two new variable $p_\pm(t,\theta)$, defined as
\be\label{eq:BFdic}
\r(t, \q) =\frac{p_+(t,\q)-p_-(t,\q)}{2\pi}, \quad  \tand \quad 
v(t,\q) = \frac{p_+(t,\q)+p_-(t,\q)}{2}.
\ee
The equations for $p_\pm(t,\theta)$ are given by,
\ben\label{eq:fermisurfeom}
\dow_t p_{\pm}(t,\q) + p_{\pm}(t,\q)\dow_\q p_{\pm}(t,\q) +W'(\q) =0.
\een
The collective Hamiltonian (\ref{eq:HamB}), written in terms of these new variables $p_\pm(t,\theta)$, is decomposed into two disjoint sectors $H_p^+$ and $H^-_p$ given by
\ben\label{eq:HBpm}
H_B = H_B^+ + H_B^-, \quad \text{where} \quad H_B^\pm = \pm \frac{1}{2\pi}\int d\theta \lb \frac{p_\pm(t,\theta)^3}{6}+W(\theta) p_\pm(t,\theta)\rb.
\een
The equations for $p_\pm$ can be derived from this Hamiltonian for the following Poisson's structure
\ben\label{eq:poissoon}
\{ p_\pm(t,\theta), p_\pm(t,\theta')\} = \mp 2\pi \delta'(\theta - \theta'), \quad  \{ p_+(t,\theta), p_-(t,\theta')\} = 0.
\een

\subsection{The droplet description}\label{sec:droplet}

Decomposition of the Hamiltonian enables us to give a geometric description of collective field theory in terms of two-dimensional droplets \cite{polchinski}. The set of decoupled equations (\ref{eq:fermisurfeom}) governs the evolution of a free-fermi droplet in $(p,\theta)$ plane whose boundaries are given by $p_\pm(t,\theta)$. To understand this in detail, we consider a system of $N$ non-relativistic free fermions (non-interacting) moving on $S^1$ under a common potential $W(\theta)$. The single particle Hamiltonian is given by
\ben\label{eq:singlep}
\mathfrak{h}(p,\theta) = \frac{p^2}2 + W(\q).
\een
The Hamilton's equations obtained from the single particle Hamiltonian (\ref{eq:singlep}) are given by
\ben\label{eq:speqn}
\frac{dp}{dt} = -W'(\theta), \qquad \frac{d\theta}{dt} = p.
\een
The single particle phase space of this system is given by a droplet with boundaries $p_\pm(t,\theta)$. Using equations (\ref{eq:speqn}) one can check that the boundaries of a droplet $p\ra p_\pm(t,\theta)$ follow equation (\ref{eq:fermisurfeom}). Therefore the equations in (\ref{eq:fermisurfeom}) determine classical evolution of the fermi surface with time. The phase space Hamiltonian for such free fermi system is given by,
\be\label{eq:HamF}
H_{\text{pp}} = \frac1{2\pi} \int d\q \ dp \lb \frac{p^2}2 + W(\q) \rb \varpi(p,\q)
\ee
where $\varpi(p,\q)$ is the phase space density
\ben
\varpi(p,\q) = \Theta\lb(p_+(t,\theta)-p)(p-p_-(t,\theta))\rb.
\een
Integrating over $p$, it is easy to check that the phase space Hamiltonian $H_{\text{pp}}$ is exactly same as the bosonic Hamiltonian given by Eqn. (\ref{eq:HBpm}).

There is a one to one correspondence between phase space variables and collective field theory variables. Eigenvalue density and the corresponding momentum can be obtained from phase space distribution by integrating over $p$
\ben \label{eq:BFdic2}
\r(t,\q) =\frac1{2\pi} \int dp \ \varpi(p, \q), \quad
v(t,\q) = \frac1{2\pi\r} \int dp \ p \ \varpi(p,\q). 
\een
Thus the relations (\ref{eq:BFdic}) serve as a dictionary between bosonic (collective field theory) and fermionic (phase space) variables.

Solving the field theory equations of motion (\ref{eq:colleom}) is equivalent to solving for upper and lower fermi surfaces in phase space picture. In either case, one needs to provide an initial data on a constant time slice in $(t,\q)$ plane. After that the problem reduces to a {\it Cauchy problem}. Existence of a unique solution depends on the geometry of initial data curve\footnote{See \cite{pallab}, for an example.}.

\subsection{Folds}\label{sec:folds}

\begin{figure}[h]
		\begin{center}
  \includegraphics[width=.6\textwidth]{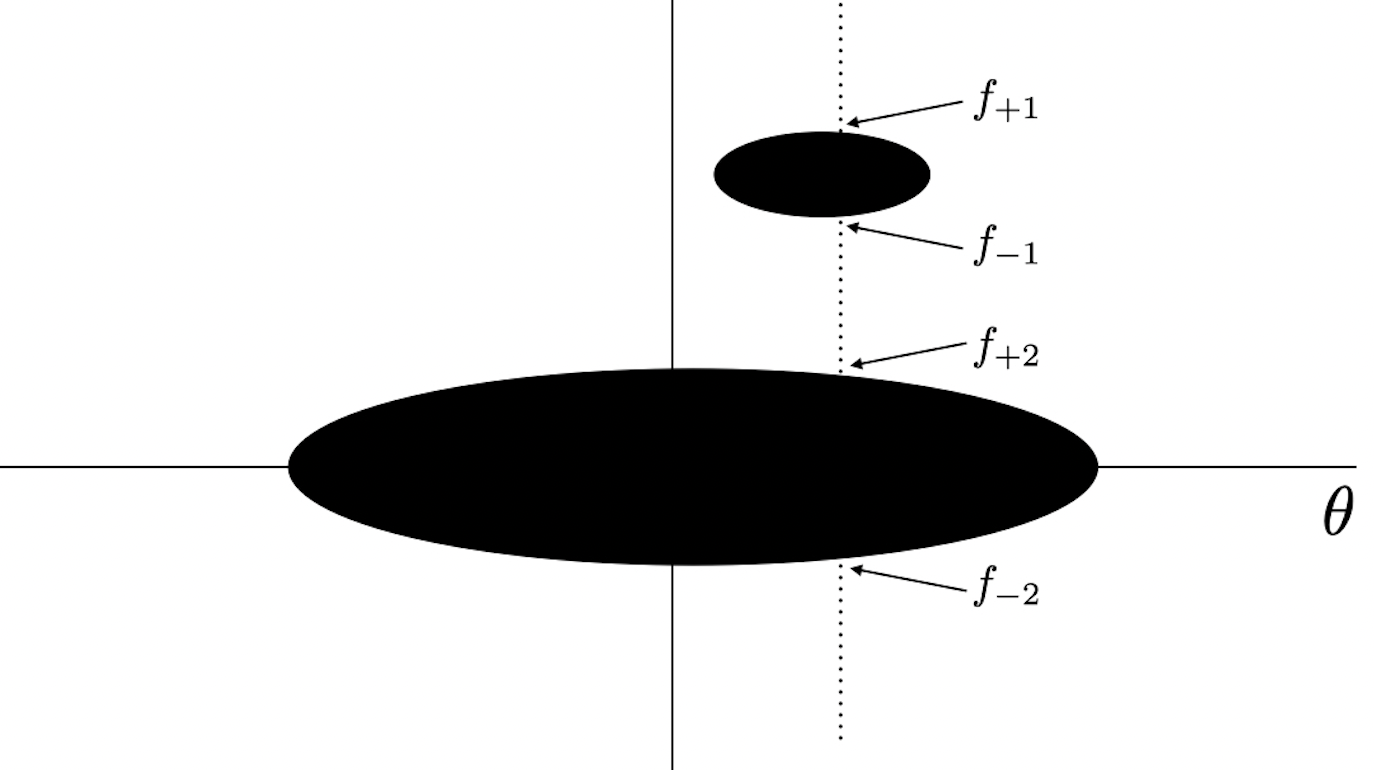}
		\end{center}
		\caption{Folds on fermi surface.}
  \label{fig:onefold}
	\end{figure}

The fermi surfaces ($p_\pm(t,\theta)$) we have discussed so far are single valued functions of $\theta$ at any time $t$. A more generic fermi surface can be multi valued in $\theta$ (see fig. \ref{fig:onefold}). Such a surface is called a folded fermi surface. The folds can appear both in the upper and the lower surfaces. The folds can be connected to the main droplet or can be a separate droplet also. The classical theory of folds were discussed in \cite{Das:1995gd, Das:2004rx} in terms of $p_\pm(t,\theta)$ and an infinite set of variables $w_{\pm n}(t,\theta)$ satisfying classical $w_\infty$ algebra \cite{Avan:1992gm}. In presence of folds one can parametrise different moments of $p$ as
\begin{align}
\label{eq:genfoldrelation1}
\int \varpi(p,\theta) & = p_+ - p_- \\
\label{eq:genfoldrelations}
    \int \frac{p^n}{n} \varpi(p,\theta) & = \frac{1}{n(n+1)} \lb p_+^{n+1} - p_-^{n+1}\rb + \sum_{k=1}^n c^n_{k} \lb p_+^{n-k}w_{+ k} - p_-^{n-k}w_{- k} \rb   
\end{align}
up to some constants $c^n_k$. These constants can be fixed from the integrable structure of the model. We shall fix them later. Presence of non-zero $w_{\pm n}$s signifies non-quadratic profile of the droplet. For example, we consider a droplet with one fold in the upper surface as shown in fig.\ref{fig:onefold}. 
For a given $\theta$ we have $\{f_{+1}, f_{-1}\}$ and $\{f_{+2}, f_{-2}\}$. From the above set of relations in (\ref{eq:genfoldrelation1}) we have
\begin{eqnarray}\label{eq:momfold0}
           \int dp \ \varpi(p,\theta) & = f_{+1}-f_{-1} + f_{+2} - f_{-2} = p_+ - p_- .
\end{eqnarray}
Since the fold is formed on the upper surface (in this particular example) it is natural to parametrise
\begin{equation}
    p_+ = f_{+1}-f_{-1} + f_{+2}, \quad \text{and} \quad p_- = f_{-2}.
\end{equation}
From the equations for higher order moments (\ref{eq:genfoldrelations}) one can find all the $w_{\pm n}$. For the fold shown in fig.\ref{fig:onefold}, first few of them are given by
\begin{eqnarray}\label{eq:wfrel}
    \begin{split}
        w_{+1} & = -\frac{1}{c^1_1}\left(f_{\text{-1}}-f_{\text{+1}}\right) \left(f_{\text{-1}}-f_{\text{+2}}\right), \\
        w_{+2} & = \frac{1}{2 c^2_2} w_{+1} \left(2c^2_1
   f_{\text{-1}} + (c^1_1 -2 c^2_1) (f_{\text{+1}}+f_{\text{+2}})\right) \\
   & \hspace{.25cm} \vdots
    \end{split}
\end{eqnarray}
and all $w_{-n} = 0$. From these relations we see that when fold is absent \emph{i.e.} $f_{+1} = f_{-1}$ or $f_{-1} = f_{+2}$ we have $p_+ = f_{+2}$, $p_- = f_{-2}$ and $w_{\pm n} =0$. In general, there can be $T$ folds on the upper surface. Then the parameterisation is given by
\begin{eqnarray}\label{eq:pfrel}
    p_+ = \sum_{i=1}^T  f_{+i} - \sum_{i=1}^{T-1} f_{-i}, \quad p_- = f_{-T}
\end{eqnarray}
and the $w_{\pm n}$ can be found as functions of $f_{\pm i}$ using (\ref{eq:genfoldrelations}):
\begin{eqnarray}\label{eq:pfrel2}
    \sum_{i=1}^T \frac{f_{+i}^{n+1} - f_{-i}^{n+1}}{n(n+1)} = \frac{1}{n(n+1)} \lb p_+^{n+1} - p_-^{n+1}\rb + \sum_{k=1}^n c^n_{k} \lb p_+^{n-k}w_{+ k} - p_-^{n-k}w_{- k} \rb.
\end{eqnarray}
The number of folds can depend on the position $\theta$ and time $t$.

In the context of 2D string theory $w_{\pm n}$ represents an infinite set of discrete fields and $p_\pm$ represent tachyon fields \cite{Avan:1992gm}. The discrete fields themselves close a classical $w_\infty$ algebra. Interpretation of $w_{\pm n}$ as folds on fermi surfaces was given by \cite{Das:1995gd}. Classical evolution of folded fermi surfaces is therefore governed by the dynamics of $p_\pm$ and $w_{\pm n}$ for a given Hamiltonian \cite{Das:1995gd}. Later we shall see how the dynamics of higher-spin fields in $AdS_3$ is mapped with the evolution of folded fermi surfaces.

\section{Higher spin gravity in  $AdS_3$ and Chern-Simons theory} \label{sec:hsgravity}

We consider gravity in $AdS_3$ coupled with integral higher spin fields with spin $3\leq s \leq M$. The bulk theory can be formulated in terms of Chern-Simons theory with the gauge group $SL(M, {R}) \times SL(M, {R})$ \cite{Campoleoni:2010zq, Campoleoni:2011hg}. The matrix valued gauge fields can be separated in two chiral sectors and denoted by $A^\pm(x)$. $'+'$ and $'-'$ correspond to gauge fields associated with two copies of the gauge groups. Two gauge fields $A^\pm(x)$ are related to generalised vielbein $e$ and spin connection $\o$ and formally expressed as \cite{Campoleoni:2010zq} (the index structure is suppressed)
\be\label{eq:gaugemetricrel}
A^\pm = \o \pm \frac{e}l
\ee
where $l$ is the radius of the $AdS_3$ space. Using the relation between the generalised vielbein, spin connection and gauge fields, the Einstein-Hilbert action can be shown to be equal to CS action
\ben\label{eq:CSacn}
I = I_{CS}(A^+) + I_{CS}(A^-)
\een
where,
\ben\label{eq:CSacn2}
I_{CS}(A^\pm) = \pm \frac{k_M}{4\pi} \int \Tr\lB A^\pm \wedge dA^\pm + \frac23 \lb A^\pm\rb^3 \rB + \cB_\infty(A^\pm).
\een
The level $k_M$ of the CS theory is related to $3$ dimensional Newton's constant $G$ and $AdS$ radius $l$ through 
\begin{eqnarray}
    k_M = \frac{l}{4G \epsilon_M} \quad \text{where} \quad  \epsilon_M = \frac{M(M^2-1)}6.
\end{eqnarray}
For $M=2$
\begin{eqnarray}\label{eq:kdef}
    k_2 = \frac{l}{4G} \equiv k.
\end{eqnarray}
$\cB_\infty(A^\pm)$ is a boundary term added to the bulk action to make $\delta I_{CS}(A^\pm) = 0$. The gauge fields $A^\pm$ are matrix valued and the trace in (\ref{eq:CSacn2}) acts on the generators of the algebra $\mathbf{sl}(M,R)$ in the fundamental representation. 

In this paper we consider the principal embedding of $\mathbf{sl}(2,R)$ in $\mathbf{sl}(M,R)$ \cite{Grumiller:2016kcp}. The generators of $\mathbf{sl}(M,R)$ are given by $L_\pm$ and $L_0$ and $W_m^{(s)}$ where $s=3, \cdots , M$ and $m = -(s-1), \cdots, (s-1)$. Important relations between these generators, needed in this paper are given by,
\begin{eqnarray}
    \begin{split}
        \Tr\lb L_0 L_0\rb = \frac{\epsilon_M}{2}, \quad 
        \Tr\lb L_0 W_0^{(s)}\rb = 0, \quad 
        \Tr \lb W_0^{(s)} W_0^{(s)}\rb = \frac{\epsilon_M}{2\sigma_s^2}
    \end{split}
\end{eqnarray}
where
\begin{equation}
    \sigma_s = \sqrt{\frac{(2s-1)! (2s-2)!}{48 (s-1)!^4 \prod_{k=2}^{s-1}(M^2 - k^2)}}.
\end{equation}
For other details see \cite{Grumiller:2016kcp}. Different components of the metric can be computed from the gauge fields
\begin{equation}\label{eq:metric}
g_{\mu\nu} = \frac{l^2}{2} \Tr\lB (A^+ - A^-)_\mu (A^+ - A^-)_\nu \rB.
\end{equation}

In 3 dimensions gravity is locally trivial, all the dynamics is localised near the boundary and hence sensitive to the boundary conditions.

\subsection{Boundary conditions}

We parametrise the three dimensional manifold by $r$, $t$ and $\theta$, where $\theta$ is compact. $r$ is the radial direction and the asymptotic region (boundary) is at $r\ra \infty$. We consider the following form of the gauge fields
\ben\label{eq:Apm-apm}
A^\pm = b_\pm^{-1} \lb d + a^\pm \rb b_\pm
\een
where $b^\pm$ are gauge group elements, they depend on the radial coordinate $r$ only. The connections $a^\pm$ depend only on the transverse coordinates $t$ and $\theta$. We must emphasize here that the specific radial dependence of $b_{\pm}$ does not affect the asymptotic charges and the boundary equations of motion. However, in order to write an explicit form of the bulk metric, one is required to make a specific choice of $b_{\pm}$. In order to satisfy the Maxwell's equations (i.e. Einstein's equation) we do not need to specify any particular form of $b_\pm$. We choose the connection in the following form \cite{Grumiller:2016kcp}
\begin{eqnarray}\label{eq:apmform}
    a^\pm(t,\theta) =  \lb \xi_\pm(t,\theta) dt \pm p_\pm(t,\theta) d\theta \rb L_0 + \sum_{s=3}^M \sigma_s \lb \zeta_{\pm s}(t,\theta) dt \pm u_{\pm s}(t,\theta) d\theta \rb W_0^{(s)}
\end{eqnarray}
where $p_\pm$ and $u_{\pm s}$ are the dynamical fields associated with gravity and higher spin fields respectively. $\xi_\pm$ and $\zeta_{\pm s}$ are corresponding chemical potentials. The Maxwell's equations obtained from the action (\ref{eq:CSacn2}) are given by
\begin{eqnarray}\label{eq:Maxeq}
    dA^\pm + A^{\pm 2} =0.
\end{eqnarray}
Using the form of $A^\pm$ and $a^\pm$ one can check that the dynamical fields and the chemical potentials satisfy
\be\label{eq:maxeq}
\Dot{p}_\pm(t,\theta) = \pm \xi_\pm'(t,\theta), \quad \dot u_{\pm s}(t,\theta) = \pm \zeta_{\pm s}'(t,\theta).
\ee
Here $\cdot$ and $'$ denote the partial derivatives with respect to $t$ and $\theta$ respectively.

In order to find the boundary term $\cB_\infty^\pm$ we demand that $\delta I = 0$ for any arbitrary variations of gauge fields. This implies
\begin{eqnarray}\label{eq:Bterms}
    \delta \cB_\infty^\pm = \mp \frac{k}{4\pi} \int dt d\theta \lb \xi_\pm \delta p_\pm + \sum_{s=3}^M \zeta_{\pm s}\delta u_{\pm s}\rb.
\end{eqnarray}
Therefore to get a well-defined boundary term one needs to take the $\delta$ outside the integral. It is possible if the chemical potential $\xi_\pm$ and $\zeta_{\pm s}$ can be written as variation of a quantity $H^\pm$ with respect to $p_\pm$ and $u_{\pm s}$ respectively, i.e.
\begin{eqnarray}\label{eq:xidef}
    \xi_\pm = - \frac{4\pi}{k} \frac{\delta H^\pm}{\delta p_\pm} \quad \text{and} \quad  \zeta_{\pm s} = - \frac{4\pi}{k} \frac{\delta H^\pm}{\delta u_{\pm s}}\ , 
\end{eqnarray}
where $H^\pm$ in general can be a functional of $p_\pm$, $u_{\pm s}$ and their different $\theta$ derivatives
\ben
H^\pm = \int d\theta \ \cH^\pm(p_\pm,\{u_{\pm s}\}).
\een
The boundary term, therefore, becomes
\begin{eqnarray}\label{eq:boundterm}
\cB_\infty^\pm  = \pm \int dt d\theta \ \cH^\pm = \pm \int dt  \ H^\pm.
\end{eqnarray}
Thus we specify the boundary conditions through the choice of the function $\cH^\pm$. The dynamics of the fields $p_\pm$ and $u_{\pm s}$ therefore depends on the choice of the boundary conditions $\cH^\pm$
\begin{eqnarray}
   \Dot{p}_\pm(t, \theta) = \mp \frac{4\pi}{k} \frac{\partial }{\partial \theta} \lb  \frac{\delta H^\pm}{\delta p_\pm} \rb, \quad \dot u_{\pm s} = \mp \frac{4\pi}{k} \frac{\partial }{\partial \theta} \lb  \frac{\delta H^\pm}{\delta u_{\pm s}} \rb.
\end{eqnarray}

\subsection{Conserved charges}\label{sec:conservedcharge}

The asymptotic symmetries are determined by a set of gauge transformations that preserve the asymptotic form of the gauge fields. The gauge transformation is given by
\begin{eqnarray}\label{eq:gaugetran}
\delta a^\pm = d \lambda^\pm + [a^\pm,\lambda^\pm]
\end{eqnarray}
where $\lambda^\pm$ are the gauge transformation parameters.
One can check that the asymptotic form of the gauge fields (\ref{eq:apmform}) are preserved under (\ref{eq:gaugetran}) with the following choice of gauge transformation parameters \cite{Grumiller:2016kcp}
\be
\lambda^\pm = \eta_\pm(t,\theta) L_0 + \sum_{s=3}^M \sigma_s \eta_{\pm s}(t,\theta) W_0^{(s)}.
\ee
With this choice the asymptotic fields $p_\pm$, $u_{\pm s}$ and chemical potentials $\xi^\pm$ transform as
\begin{align}\label{eq:pgt}
    \delta p_\pm & = \pm \eta'_{\pm}, \quad \text{and} \quad \delta \xi_\pm = \Dot{\eta}_\pm \\
    \label{eq:hgt}
    \delta u_{\pm s} & = \pm \eta_{\pm s}' \quad \text{and} \quad \delta \zeta_{\pm s} = \dot{\eta}_{\pm s}. 
\end{align}
Since the chemical potentials $\xi_\pm$ and $\zeta_{\pm s}$ now depend on $p_\pm$ and $u_{\pm s}$, their variations are not zero anymore. Hence we can use (\ref{eq:xidef}) to write
\begin{align}\label{eq:etaeqn}
\dot \eta_\pm(t,\theta) = \mp \frac{4\pi}{k} \frac{\d}{\d p_\pm(t,\theta)} \int d\theta' \lb \frac{\d H^\pm}{\d p_\pm} \eta'_\pm(t, \theta') + \sum_{s=3}^M \frac{\d H^\pm}{\d u_{\pm s}} \eta'_{\pm s}(t, \theta') \rb,\\
\label{eq:etaeqn2}
\dot \eta_{\pm s}(t,\theta) = \mp \frac{4\pi}{k}\frac{\d}{\d u_{\pm s}(t,\theta)} \int d\theta' \lb \frac{\d H^\pm}{\d p_\pm} \eta'_\pm(t, \theta') + \sum_{s=3}^M \frac{\d H^\pm}{\d u_{\pm s}} \eta'_{\pm s}(t, \theta') \rb.
\end{align}
The variation of the conserved charges for the local gauge symmetry is given by \cite{REGGE1974286, Banados:1994tn}
\begin{eqnarray}
\begin{split}
    \delta \mathbf{Q}^\pm(\eta_\pm, \eta_{\pm s}) & = -\frac{k_M}{2\pi} \int d\theta' \Tr\lb \lambda^\pm \delta (a^\pm_{\theta})\rb\\
    & = \mp \frac{k}{4\pi} \int d\theta' \lb \eta_\pm \delta p_\pm + \sum_{s=3}^M \eta_{\pm s} \delta u_{\pm s}\rb.
\end{split}
\end{eqnarray}
First we can check that if $\eta_\pm = - \frac{4\pi}{k}\frac{\partial H^\pm}{\partial p_\pm}$ and $\eta_{\pm s} = - \frac{4\pi}{k}\frac{\partial H^\pm}{\partial u_{\pm s}}$ then eqns. (\ref{eq:etaeqn}) and (\ref{eq:etaeqn2}) are satisfied and hence 
\ben\label{eq:conham}
\mathbf{Q}^\pm(\eta_\pm, \eta_{\pm s}) = \pm \int d\theta \cH^\pm
\een
are conserved charges, which are the Hamiltonians for the two chiral sectors. Moreover, one can also find some other $H^\pm_n = \int d\theta \cH^\pm_n$ such that 
\begin{equation}\label{eq:etachoice}
\eta_\pm = - \frac{4\pi}{k} \frac{\partial H^\pm_n}{\partial p_\pm} \quad \text{and} \quad \eta_{\pm s} = - \frac{4\pi}{k} \frac{\partial H^\pm_n}{\partial u_{\pm s}}
\end{equation}
satisfy eqns. (\ref{eq:etaeqn}) and (\ref{eq:etaeqn2}) for a given choice of Hamiltonians $H^\pm$. In that case 
\ben\label{eq:concharge}
\mathbf{Q}^\pm_n(\eta_\pm, \eta_{\pm s}) = \pm \int d\theta \ \cH^\pm_n
\een
are also a conserved charges of the theory under the boundary conditions specified through $H^\pm$. Apparently the right hand side does not depend on $\eta_\pm$ and $\eta_{\pm s}$, however it depends on the choice of these quantities.

Poisson brackets of charges satisfy \cite{REGGE1974286}
\begin{eqnarray}
    \delta_{\{\tilde\eta_\pm, \tilde\eta_{\pm s} \}} \mathbf{Q}^\pm_n (\eta_\pm, \eta_{\pm s}) = \{ \mathbf{Q}^\pm_n (\eta_\pm, \eta_{\pm s}), \mathbf{Q}^\pm_n (\tilde\eta_\pm, \tilde\eta_{\pm s})\}.
\end{eqnarray}
From these relations we can deduce the Poisson brackets for the field $p_\pm$ and $u_{\pm s}$
\begin{align}\label{eq:pPB}
    \{p_\pm(t,\theta), p_\pm(t, \theta')\} & = \mp \frac{4\pi}{k} \frac{\partial}{\partial \theta}\delta(\theta-\theta'), \\
    \label{eq:usPB}
    \{u_{\pm s}(t,\theta),u_{\pm s'}(t, \theta')\} & = \mp \frac{4\pi}{k} \frac{\partial}{\partial \theta}\delta(\theta-\theta') \delta_{ss'}\\
    \label{eq:puPB}
   \{p_{\pm}(t,\theta),u_{\pm s'}(t, \theta')\} & = 0.
\end{align}
Once we have the Poisson brackets of the field variables the field equations (\ref{eq:maxeq}) can be written as
\begin{eqnarray}
    \dot p_\pm(t,\theta) = \{p_\pm(t,\theta), H^\pm\}, \quad \text{and} \quad  \dot u_{\pm s}(t,\theta) = \{u_{\pm s}(t,\theta), H^\pm \}.
\end{eqnarray}

In \cite{Ojeda:2019xih, Ojeda:2020bgz}, $\cH_n$s are chosen to be generalised Gelfand-Dikii polynomials (or modified Gelfand-Dikii polynomials for modified Boussinesq hierarchy in presence of higher spins) and the field equations are give by left and right members of Gardner hierarchy (or modified Boussinesq hierarchy for higher spin). In the next section we shall see that a different set of $\cH_n$s is possible and such choices will provide a gravity dual of integrable collective field theory.

\section{Collective field theory and $AdS_3$ gravity}\label{sec:CFTgravity}

In this section we discuss the gravity dual of interacting collective field theory. Our construction allows us to provide a geometric description of the bulk solution in terms of the shapes of fermi surfaces. We first do the exercise for pure gravity and show how the boundary dynamics is captured by the time evolution of free fermi droplets. After that we discuss the higher spin case.

To match the classical dynamics of gravity and that of collective field theory given by eqns. (\ref{eq:maxeq}) and (\ref{eq:fermisurfeom}) respectively we take the Hamiltonian $H^\pm$ to be proportional to the cubic collective field theory Hamiltonian (\ref{eq:HBpm})
\begin{equation}\label{eq:HpmHBB}
    H^\pm = \frac{k}{2}H^\pm_{B} = \pm \frac{k}{4\pi} \int d\theta dp \ \mathfrak{h}(p,\theta) \varpi(\theta, p).
\end{equation}
The equations of motion satisfied by $p_\pm$ are given by (\ref{eq:fermisurfeom}). These are dispersion-less KdV equations with the source term\footnote{Our boundary conditions are related to the standard boundary conditions \cite{Perez:2016vqo} by a gauge transformation on-shell \cite{Afshar:2016wfy}.}. Further, we can check that for this choice of Hamiltonian, $\eta_\pm \sim \partial H^\pm/\partial p_\pm$ satisfy eqn. (\ref{eq:etaeqn}) and hence the Hamiltonian is a conserved quantity. 

Thus we introduce the new boundary conditions for $AdS_3$ gravity by choosing the boundary Hamiltonian $H^\pm$ in (\ref{eq:xidef}) to be proportional to the collective field theory Hamiltonian and hence the dynamics of boundary gravitons is captured by that of non-interacting non-relativistic fermions.

These boundary conditions also admit infinite sequences of conserved charges in the theory.  Following  \cite{Avan:1992hv, Avan:1991kq, Dhar:2005qh} we can construct an infinite sequence of phase space integrals of different integer powers of the single particle Hamiltonian $\mathfrak{h}(p,\theta)$ (given by (\ref{eq:singlep}))
\ben
H_{2n-1} = \frac{k}{4\pi} \int d\theta dp\ \varpi(p, \q)  \ \mathfrak{h}^n(p,\theta) \equiv \int d\theta \ \cH_{2n-1}(\theta) .
\een
Separating these conserved charges into two chiral sectors for each $n$ we have
\ben
H_{2n-1} = H_{2n-1}^+ + H_{2n-1}^-, \quad \text{where} \quad H_{2n-1}^\pm = \pm\frac{k}{4\pi} \int d\theta \int_0^{p_\pm} dp \lb \frac{p^2}{2} + W(\theta) \rb^n
\een
where $H_1^\pm \equiv H^\pm$ is the Hamiltonian. Expanding the power on the right hand side and integrating over $p$, one can write
\ben\label{eq:conservecharge}
H_{2n-1}^\pm = \pm \frac{k}{4\pi} \int d\theta \sum_{k=0}^n\lb \frac{\ ^nC_k }{2^k (2k+1)} p_\pm^{2k+1} W(\theta)^{n-k}\rb .
\een
It turns out that for these $H_{2n-1}^\pm$, the gauge transformation parameters $\eta_\pm$ given by (\ref{eq:etachoice}), satisfy eqn. (\ref{eq:etaeqn}) for the Hamiltonian given by eqn. (\ref{eq:HpmHBB}) for all $n$. Thus the phase space integrals (\ref{eq:concharge}) provide two infinite sequences (for two chiral sectors) of conserved charges of the $AdS_3$ gravity. Using (\ref{eq:pPB}) we see that the Poisson brackets of charges $H_{2n-1}$ for all $n>2$ with the Hamiltonian vanish and hence they are constants of motion. One can also construct the conserved current associated with these charges. We define a current density  $J^{\mu}_{(n)\pm} = \{\cH_{n}^\pm, J^\theta_{(n)\pm} \}$ such that
\ben
\del_\mu J^{\mu}_{(n)\pm} = 0 \quad \text{on-shell}.
\een
It turns out that to satisfy the on-shell conservation equation the spatial component of the current is given by,
\be
J^\theta_{(n)\pm}(\theta) = \pm  \frac{k}{4\pi(n+1)}\lb \frac{p_\pm^2}{2} + W(\theta)\rb^{n+1}.
\ee
Given the Poisson structure (\ref{eq:pPB}), one can show that the Poisson brackets of any two conserved charges vanishes
\begin{equation}
    \{\mathbf{Q}_n^\pm, \mathbf{Q}_m^\pm\} =0\quad \forall \ m,n \geq 1
\end{equation}
implying the integrable structure of the $AdS_3$ gravity for the chosen boundary conditions.

One can also take the boundary Hamiltonian to be $H_{2n-1}^\pm$ for any fixed $n>1$ to specify the boundary conditions. In that case the equations of motion become
\begin{equation}
    \dot{p_\pm} + n \mathfrak{h}(p_\pm,\theta)^{n-1} \lb p_\pm p_\pm' + W' \rb = 0.
\end{equation}
These are the set of hierarchical equations with respect to (\ref{eq:fermisurfeom}). For this choice the other $H^\pm_{2n-1}$s turn out to be the conserved charges of the motion. This shows the hierarchical nature of the $AdS_3$ gravity under the boundary conditions we considered.

\subsection{BTZ black holes and droplets}\label{sec:btz}

\begin{figure}[h]
		\begin{center}
  \includegraphics[width=.8\textwidth]{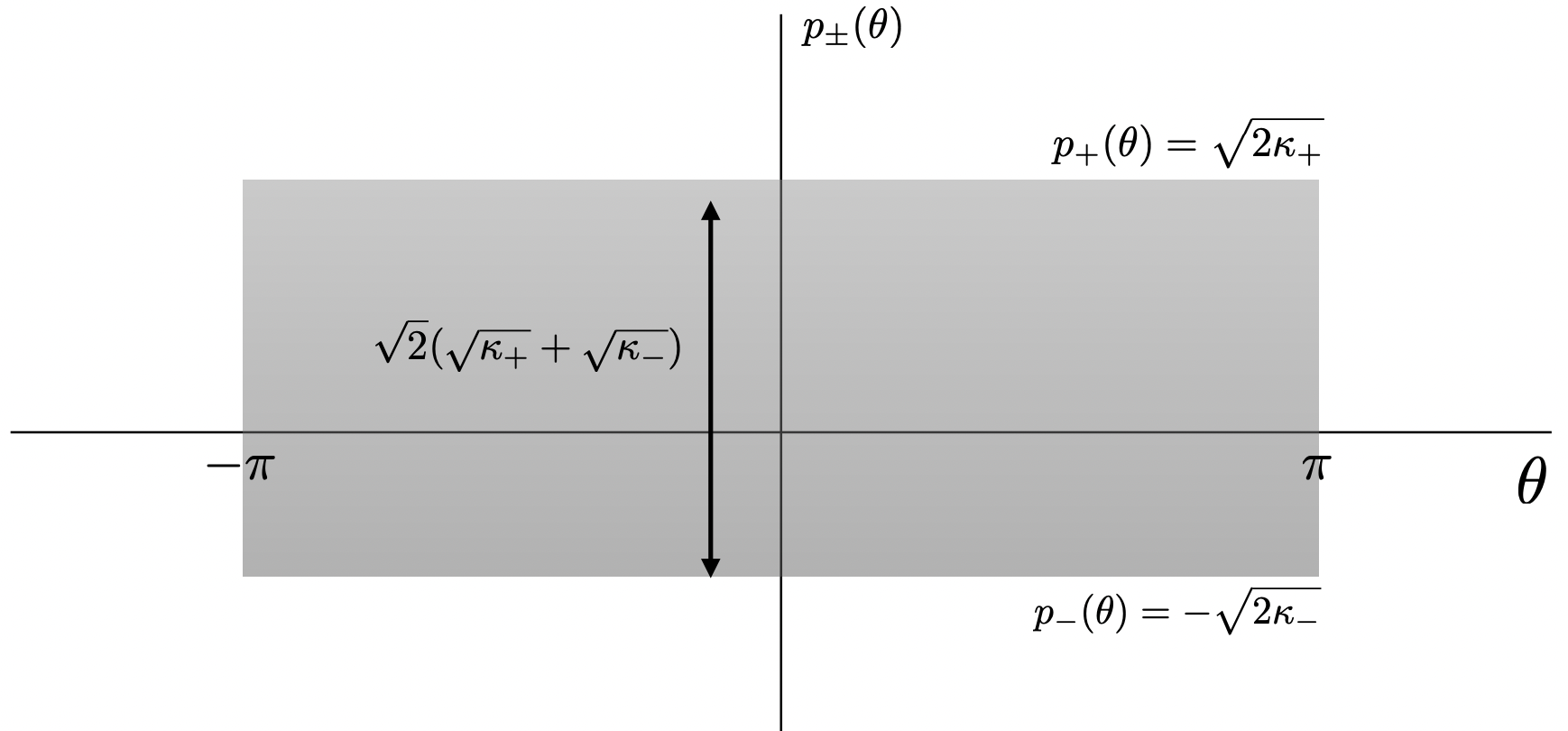}
		\end{center}
		\caption{Droplet for BTZ black hole.}
  \label{fig:constantdroplet}
	\end{figure}

In case of pure gravity the dynamical equations (\ref{eq:fermisurfeom}) admit time independent solutions. For $\dot p_\pm =0$ we have $\xi^{\pm '} =0$. This implies $\xi^{\pm}$ are functions of time only. Since we are interested in time independent solutions we consider $\xi^{\pm} = c_{\pm} (\text{constant})$ and hence we get $p_\pm(\theta) = \pm \sqrt 2 \sqrt{c_\pm - W(\theta)}$. The metric is given by
\begin{equation}
\begin{split}
   ds^2 = {dr}^2 & + \frac{l^2}{4} \cosh ^2\left(\frac{r}{l}\right)
   \left(\left(c_+-c_-\right) \text{dt}+\text{d$\theta
   $} \left(p_-(\theta )+p_+(\theta )\right)\right){}^2\\
   & -\frac{l^2}{4} \sinh ^2\left(\frac{r}{l}\right)
   \left(\left(c_-+c_+\right) \text{dt}+\text{d$\theta
   $} \left(p_+(\theta )-p_-(\theta )\right)\right){}^2.
   \end{split}
\end{equation}
The explicit form of the above metric is written using the maps \eqref{eq:metric}, \eqref{eq:Apm-apm} and \eqref{eq:apmform} where the gauge group element $b_{\pm}$ is given by
\begin{equation}
    b_{\pm}= \exp \left[\pm \frac{r}{2l}(L_{+1}-L_{-1}) \right]\ .
\end{equation}
Such solutions are called black flower solutions \cite{Grumiller:2019tyl, Grumiller:2019fmp, Afshar:2016kjj}. Since the eigenvalue density is given by $\rho(\theta) = (p_+(\theta) - p_-(\theta))/2\pi$, such solutions correspond to gapped or no-gap solution in the matrix model side depending whether $p_{\pm}(\theta)$ is defined over the whole range of $\theta$ or not. A further special case is constant solution : $p_\pm(\theta) = \pm \sqrt{2} \sqrt{\kappa_\pm}$. After a suitable coordinate transformation
\begin{equation}
\begin{split}
    r & = \frac{l}{2} \log \left[\frac{\left(\kappa _-+\kappa _+\right) l^2 - 2 \tilde{r}^2 -\sqrt{ 4 \tilde{r}^4-4 \left(\kappa _-+\kappa _+\right) l^2 \tilde{r}^2  + \left(\kappa _+-\kappa
   _-\right){}^2 l^4} }{2 \sqrt{\kappa _- \kappa _+} l^2}\right]\\
   t & = \frac{2 \sqrt{2} \sqrt{\kappa _+ \kappa _-} v}{l \left(c_- \sqrt{\kappa _+}-c_+ \sqrt{\kappa _-}\right)}, \quad
   \theta  = \phi  + \frac{c_+ \sqrt{\kappa _-} + c_- \sqrt{\kappa _+}}{l \left(c_+ \sqrt{\kappa _-} - c_- \sqrt{\kappa _+}\right)}v
   \end{split}
\end{equation}
the metric can be written in the standard Schwarzschild coordinate,
\begin{equation}
    ds^2 = -f(\tilde r)dv^2 + \frac{d\tilde r^2}{f(\tilde r)} + \tilde r^2 \left( d\phi - \frac{J}{2\tilde r^2} dv\right)^2.
\end{equation}
The function $f(\tilde r)$ is given by
\begin{equation}
    \begin{split}
       f(\tilde r) & = \frac{\tilde r^2}{l^2} -M + \frac{J^2}{4\tilde r^2} \\
    \text{where}\quad M & = (\kappa_+ + \kappa_-)\\
      \text{and}\quad J & =  l (\kappa_+ - \kappa_-).
    \end{split}
\end{equation}
The solution has horizons at
\ben
r_{\pm} = \frac{l}{\sqrt 2} \lb \sqrt{\kappa_+} \pm \sqrt{\kappa_-}\rb.
\een
The entropy is given by
\ben
S_{BH} = \frac{2\pi r_+}{4G} = \sqrt{2}\pi k \lb \sqrt{\kappa_+} + \sqrt{\kappa_-}\rb.
\een
The constant configuration $p_\pm(\theta) = \pm \sqrt{2 \kappa_\pm}$ corresponds to droplet as shown in fig.\ref{fig:constantdroplet}.
The area of the droplet is given by
\ben
\cA = 2\pi (p_+(\theta) - p_-(\theta)) = 2\sqrt{2}\pi (\sqrt{\kappa_+} + \sqrt{\kappa_-}) = \frac{2}{k}S_{BH}.
\een
The black hole mass $M$ and angular momentum $J$ are specified in terms of droplet data $\kappa_\pm$. A symmetric distribution about $\theta$ axis (i.e. $\kappa_+ = \kappa_-$) corresponds to $J=0$, zero angular momentum. The extremal black hole corresponds to $\kappa_- = 0$. Therefore any droplet with $p_\pm(\theta) = \text{constant}$ with $|p_+| \geq |p_-|$ corresponds to a BTZ black hole.

\subsection{Higher spin droplets and conserved charges}\label{sec:higher-spin-droplet}

Since the classical algebra (Poisson brackets) satisfied by $p_\pm$ and $u_{\pm s}$ is a disjoint unions of $M-1$ Poisson brackets, a trivial generalisation of the above exercise is to construct a Hamiltonian which is a direct sum of $M-1$ copies of mutually non-interacting Hamiltonians (\ref{eq:HpmHBB}). As a result one can construct $M-1$ copies of asymptotic conserved charges (\ref{eq:conservecharge}). The boundary dynamics, similarly, can be described by $M-1$ copies of free fermi droplets. Such a generalisation is not interesting. One can, in fact, find a much more interesting boundary dynamics where higher spin excitations are coupled with spin two fields in a non-trivial way and hence conserved charges.

To specify non-trivial boundary conditions we first turn on all the integer higher spin fields and define a set of infinite number of collective variables $w_{\pm n}$ from $u_{\pm s}$
\begin{eqnarray}\label{eq:wurel}
    w_{\pm n}(t, \theta) = \sum_{s=3}^{M=\infty} \frac{u_{\pm s}^{n+1}(t, \theta)}{n+1}.
\end{eqnarray}
Using (\ref{eq:puPB}) we see that $w_{\pm m}$ satisfy the classical $w_\infty$ algebra
\begin{equation}\label{eq:wpoisson}
    \{w_{\pm m}(t,\theta), w_{\pm n}(t, \theta') \} = \mp \frac{4\pi}{k} \lb m w_{\pm m+n-1}(t,\theta) + n w_{\pm m+n-1}(t,\theta')\rb \frac{\partial}{\partial\theta} \delta(\theta - \theta').
\end{equation}

In order to specify the boundary conditions in presence of higher spin excitations we again integrate the single particle Hamiltonian $\mathfrak{h}(\theta, p)$ over phase space with folds. The boundary Hamiltonian, therefore, is given by
\begin{eqnarray}\label{eq:foldH}
\begin{split}
    H & = \frac{k}{4\pi} \int d\theta dp \varpi(\theta, p) \mathfrak{h}(\theta, p) \\
    & = \frac{k}{4\pi} \int d\theta \lB \lb \frac{p_+^3}{6} - \frac{p_-^3}{6}\rb + \lb p_+ w_{+1} - p_- w_{-1}\rb + \lb w_{+2} - w_{-2}\rb + W(\theta)\lb p_+ - p_- \rb \rB.
\end{split}
\end{eqnarray}
Here we have chosen 
\begin{eqnarray}\label{eq:choicecmn}
   c^2_1 =1 \quad \text{and} \quad c^2_2 = 1. 
\end{eqnarray}
The equations of motion satisfied by $p_\pm$ and different higher spin fields are given by
\begin{eqnarray}\label{eq:puseqn}
    \begin{split}
        \dot p_\pm +  p_\pm p_\pm' + \sum_{s=3}^\infty u_{\pm s} u_{\pm s}' + W' & = 0 \\
       \dot u_{\pm s} + p'_\pm u_{\pm s} + p_\pm u'_{\pm s} + 2 u_{\pm s}u'_{\pm s} & = 0.
    \end{split}
\end{eqnarray}
In terms of of collective excitations $w_{\pm n}$ these equations can be written as,
\begin{eqnarray}\label{eq:pweqn}
    \dot p_\pm +  p_\pm p_\pm' + w_{\pm 1}' + W' = 0 \ \ \text{and} \ \ \dot w_{\pm n} +  (n+1) p_{\pm}' w_{\pm n} + p_\pm w_{\pm n}' + 2 w_{\pm (n+1)}' = 0. \ \ 
\end{eqnarray}
Considering $p_\pm$ and $w_{\pm n}$ to be independent fields, these equations can also be obtained using the Poisson brackets of $w_{\pm n}$ (\ref{eq:wpoisson}) for the Hamiltonian (\ref{eq:foldH}). One can check that the eqns. (\ref{eq:etaeqn}) and (\ref{eq:etaeqn2}) are satisfied for the choice : $\eta_\pm \sim \frac{\partial H}{\partial p_{\pm}}$ and $\eta_{s\pm} \sim \frac{\partial H}{\partial u_{s \pm}}$. Therefore the Hamiltonian (\ref{eq:foldH}) provides consistent boundary conditions for higher spin gravity. In presence of arbitrary higher spins the above sets of equations (\ref{eq:pweqn}), in general, neither admit any integrable structure nor a free fermi description.

\subsubsection{Integrable structure} 

It turns out that if we turn on a single higher spin (spin $s=3$) i.e. 
\begin{eqnarray}\label{eq:wnnosum}
    w_{n\pm} = \frac{u_{\pm s}^{n+1}}{n+1} \quad \text{(no sum)},
\end{eqnarray}
then it is possible to construct an infinite sequence of conserved charges for each chiral sector. The conserved charges are obtained by integrating $\mathfrak{h}^n$ over phase space in presence of folds. 
\begin{eqnarray}\label{eq:nonrelcon}
    H_{2n-1} = \frac{k}{4\pi}\int d\theta dp \mathfrak{h}^n(\theta, p) = H^+_{2n-1} + H^-_{2n-1}, \quad n \geq 1.
\end{eqnarray}
The integrals are determined up to the constants $c^n_m$, defined in (\ref{eq:genfoldrelation1}) and (\ref{eq:genfoldrelations}). It turns out that one we can fix this constants (for a given choice of $c_1^2$ and $c_2^2$ in eqn.(\ref{eq:choicecmn})) by taking $\eta_\pm \sim \frac{\del H_{2n-1}^\pm}{\del p_\pm}$ and $\eta_{\pm s} \sim \frac{\del H_{2n-1}^\pm}{\del u_{\pm s}}$ such that eqns. (\ref{eq:etaeqn}) and (\ref{eq:etaeqn2}) are satisfied. We fix the coefficients $c^{2n}_1, \cdots, c^{2n}_{2n}$ to find the conserved charge $H_{2n-1}^\pm$. Thus $H_{2n-1}^\pm$ generate two infinite sequences of conserved charges of the higher spin (spin 3) gravity in $AdS_3$ with the boundary conditions specified through the Hamiltonian given in (\ref{eq:foldH}). The first few 
conserved charges (for $n=2$, $n=3$ and $n=4$) are given by (for $w_{\pm n}$ given by eqn. (\ref{eq:wnnosum}))
\begin{eqnarray}
\begin{split}
    H_3^\pm & = \pm \frac{k}{4\pi}\int d\theta \Bigg[ \frac{p_\pm^5}{20} +p_\pm^3 w_{\pm 1} + 3 p_\pm^2 w_{\pm 2} + 5 p_\pm w_{\pm 3} + 3 w_{\pm 4} \\
    & \hspace{2.5cm} + 2 W(\theta) \lb \frac{p_\pm^3}{6}  + p_\pm w_{\pm 1} +  w_{\pm 2} \rb  + W^2(\theta) p_\pm \Bigg],   
\end{split}
\end{eqnarray}
\begin{eqnarray}
\begin{split}
     H_5^\pm & = \pm \frac{k}{4\pi}\int d\theta \Bigg[ \frac{p_\pm^7}{56} + \frac{3}{4} p_\pm^5 w_{\pm 1} + \frac{15}{4} p_\pm^4 w_{\pm 2} + \frac{25}{2} p_\pm^3 w_{\pm 3} + \frac{45}{2} p_\pm^2 w_{\pm 4} + \frac{87}{4} p_\pm w_{\pm 5} \\
    & \hspace{2.5cm} + \frac{35}{4} w_{\pm 6}  + 3W(\theta) \lb \frac{p_\pm^5}{20} +p_\pm^3 w_{\pm 1} + 3 p_\pm^2 w_{\pm 2} + 5 p_\pm w_{\pm 3} + 3 w_{\pm 4} \rb \\
    & \hspace{2.5cm} + 3W^2(\theta) \lb \frac{p_\pm^3}{6}  + p_\pm w_{\pm 1} +  w_{\pm 2} \rb + W^3(\theta) p_\pm \Bigg],
    \end{split}
\end{eqnarray}
\begin{eqnarray}
\begin{split}
    H_7^\pm & = \pm \frac{k}{4\pi}\int d\theta \Bigg[ \frac{p_\pm^9}{144} + \frac{1}{2} p_\pm^7 w_{\pm 1} + \frac{7}{2} p_\pm^6 w_{\pm 2} + \frac{35}{2} p_\pm^5 w_{\pm 3} + \frac{105}{2} p_\pm^4 w_{\pm 4} + \frac{203}{2} p_\pm^3 w_{\pm 5} \\ &  + \frac{245}{2}p_\pm^2 w_{\pm 6}+ \frac{169}{2}p_\pm w_{\pm 7}+ \frac{51}{2}p_\pm^2 w_{\pm 8} + 3 W(\theta) \bigg( \frac{p_\pm^{7}}{42}+p_{\pm}^{5}w_{\pm 1} +5 p_{\pm}^{4}w_{\pm 2} \\ 
    & +\frac{50}{3}p_{\pm}^{3}w_{\pm 3} +30 p_{\pm}^{2}w_{\pm 4}+29 p_{\pm}w_{\pm 5}+\frac{35}{3}w_{\pm 6}\bigg) + 6 W(\theta)^{2} \bigg( \frac{p_{\pm}^{5}}{20}+p_{\pm}^{3} w_{\pm 1}+3 p_{\pm}^{2} w_{\pm 2} \\
    & + 5 p_{\pm} w_{\pm 3}+3  w_{\pm 4}\bigg) + 4 W(\theta)^{3}\lb\frac{p_{\pm}^{3}}{6}+p_{\pm} w_{\pm 1}+w_{\pm 2}\rb \Bigg].
\end{split}
\end{eqnarray}
Using (\ref{eq:pPB}, \ref{eq:usPB} and \ref{eq:puPB}) we find that the Poisson brackets of $H_{3}$, $H_{5}$ \emph{etc.}  with the Hamiltonian (\ref{eq:foldH}) vanish. Thus we find two infinite sets of conserved charges and hence an integrable structure for higher spin gravity.

We can also find a different set of conserved charges if we consider that the boundary dynamics of gravity and higher spin fields is governed by a system of relativistic fermions with single particle Hamiltonian $\mathfrak{h} = p + V(\theta)$. The total Hamiltonian is therefore given by
\begin{align}
     \tilde{H} & = \frac{k}{4\pi} \int d\theta dp (p + W(\theta)) \varpi(\theta,p) \nonumber \\ 
     & = \frac{k}{4\pi} \int d\theta \lB \frac{p_+^2}{2} + w_{+1} - \frac{p_-^2}{2} - w_{-1} + W(\theta)(p_+ - p_-)\rB .  \end{align}
Here again we have chosen the constant $c_1^1 = 1$. The equation of motion for this Hamiltonian is given by
\begin{eqnarray}
    \dot p_\pm =  p'_\pm + W', \qquad \dot u_{\pm s} = u'_{\pm s} .
\end{eqnarray}
For the relativistic Hamiltonian the infinite set of conserved charges are defined by
\begin{equation}\label{eq:relcon}
    \tilde H_{n} = \frac{k}{4\pi}\int d\theta dp (p+ W(\theta))^n = \tilde H^+_{n} + \tilde H^-_{n}.
\end{equation}
As before we fix the set of coefficients $c^{n+1}_1, \cdots c^{n+1}_{n+1}$ by demanding that eqns. (\ref{eq:etaeqn}) and (\ref{eq:etaeqn2}) are satisfied for $\eta_\pm \sim \frac{\del \tilde H_{n}^\pm}{\del p_\pm}$ and $\eta_{\pm s} \sim \frac{\del \tilde H_{n}^\pm}{\del u_{\pm s}}$. The first few such charges are given by\footnote{Some of the $c^{n+1}_{m}$ that remain unfixed after satisfying (\ref{eq:etaeqn}) and (\ref{eq:etaeqn2}) have been fixed by demanding $\{H_n, H_m\} =0$.}
\begin{eqnarray}
    \tilde H_2^\pm  =  \pm \frac{k}{4\pi} \int d\theta \lB \frac{p_\pm^3}{3} + 2 p_\pm w_{\pm 1} + 2 w_{\pm 2} + 2W\lb \frac{p_\pm^2}{2} + w_{\pm 1}\rb + W^2 p_\pm \rB,
\end{eqnarray}
\begin{eqnarray}
    \tilde H_3^\pm & = & \pm \frac{k}{4\pi} \int d\theta \Bigg[ \frac{p_\pm^4}{4} + 3 p_\pm^2 w_{\pm 1} + 6 p_\pm w_{\pm 2} + 5  w_{\pm 3} + 6 W \lb \frac{p_\pm^3}{6} + p_\pm w_{\pm 1} + w_{\pm 2} \rb \nonumber \\
    && \hspace{2cm}  + 3 W^2 \lb \frac{p_\pm^2}{2} + w_{+1} \rb + W^3 p_\pm \Bigg],
\end{eqnarray}
\begin{eqnarray}
\begin{split}
    \tilde H_4^\pm & = \pm \frac{k}{4\pi} \int d\theta \Bigg[ \frac{p_\pm^5}{5} +  4p_\pm^3 w_{\pm 1} + 12 p_\pm^{2} w_{\pm 2} + 20 p_\pm w_{\pm 3}+12 w_{\pm 4}\\
    & \hspace{2cm} + 12 W \lb \frac{p_\pm^4}{12} + p_\pm^{2} w_{\pm 1} + 2 p_\pm w_{\pm 2} +\frac{5}{3} w_{\pm 3} \rb  \\
    & \hspace{2cm} + 12 W^2 \lb \frac{p_\pm^3}{6} +p_\pm w_{\pm 1}+ w_{+2} \rb + 4 W^3 \lb \frac{p_\pm^{2}}{2}+w_{\pm 1}\rb +W^{4}p_\pm \Bigg].
\end{split}
\end{eqnarray}
One interesting point to note here is that for constant (or zero) potential both the sectors merge together and provide a larger class of conserved charges.

\subsubsection{Free fermi description} 
There are two possible cases. First we consider that only a single higher spin (i.e. spin 3) is turned on such that the integrable structure is preserved. In the second case we give up the integrable structure and turn on all possible higher spins. In both the cases it is possible to give a free fermi description of the $AdS_3$ solution. 

In presence of single higher spin the equations of motion are given by (\ref{eq:puseqn}). Suppose the dynamics is governed by folded fermi surfaces with boundaries $f_{\pm i}$. The fold boundaries $f_{\pm i}$ satisfy
\begin{eqnarray}\label{eq:foldeqns}
    \dot{f}_{\pm i} + f_{\pm i}f_{\pm i}' + W'(\theta) =0
\end{eqnarray}
for non-relativistic fermions. These equations are obtained from the single particle Hamiltonian (\ref{eq:singlep}). 

The phase space Hamiltonian in presence of folds is given by
\begin{eqnarray}
    \begin{split}
        H & = \frac{k}{4\pi} \int d\theta \varphi(\theta,p) \mathfrak{h}(\theta,p) = \frac{k}{4\pi} \sum_i \lB \frac{1}{6} \lb f_{+i}^3 - f_{-i}^3\rb + W'(\theta) \lb f_{+i} - f_{-i} \rb \rB.
    \end{split}
\end{eqnarray}
Equating this with the collective field theory Hamiltonian we get,
\begin{eqnarray}\label{eq:pfrels3}
\begin{split}
   \sum_{i} (f_{+i} - f_{-i}) & =  p_+ - p_- \\ 
    \sum_{i} \frac{1}{6} \lb f_{+i}^3 - f_{-i}^3\rb & = \frac{1}{6}\lb p_+^3 - p_-^3 \rb + \frac{u_{+s}^2}{2}p_+ - \frac{u_{-s}^2}{2}p_- + \frac{1}{3} \lb u_{+s}^3 - u_{-s}^3\rb.
    \end{split}
\end{eqnarray}
However, in order to make the fold equations (\ref{eq:foldeqns}) consistent with the field equations (\ref{eq:puseqn}), fold variables $f_{\pm i}$ satisfy some further constraints given by eqn. (\ref{eq:pfrel2}) with the same $c^n_k$, obtained for the calculations of conserved charges. This is little surprising that the $c^n_k$ were determined by choosing $H_n$ such that the gauge transformation parameters satisfy the corresponding equations. Such conditions have a priori no connection with the consistency between the fold equations and the higher spin field equations. The dynamics of higher spin gravity is captured by the evolution of folded fermi surfaces. Therefore different geometries or shapes of droplets (with folds) correspond to different solutions of higher spin gravity.  However, for a given higher spin solution the droplets/folds are highly constrained because of (\ref{eq:pfrel2}).

If we give up the integrable structure and turn on all the higher spins, then it is possible to parametrise eqns (\ref{eq:pfrel2}) in a different way such that the fold equations (\ref{eq:foldeqns}) and equations for $w_{\pm n}$ are consistent. However, this imposes an infinite sequence of restrictions on $w_{\pm n}$ depending on the choice of parametrisation. 

\section{Summary and Discussion}\label{sec:discussion}

\paragraph{Summary :} We introduce new sets of boundary conditions (following the work on generalised soft hairy boundary conditions \cite{Ojeda:2019xih, Grumiller:2019tyl}) for higher spin gravity in $AdS_3$ where the boundary dynamics of spin two and other higher spin fields are governed by the interacting collective field theory Hamiltonian \cite{Avan:1992gm}. However there is a difference between this Hamiltonian and the Hamiltonian introduced by Avan and Jevicki. In our case the collective field theory Hamiltonian is defined over a cylinder (\emph{i.e.} the spatial direction is a circle) and hence can be obtained from a unitary matrix quantum mechanics in presence of an arbitrary potential $W(\theta)$. The interaction of this free Hamiltonian with the $w_\infty$ excitations was introduced in \cite{Avan:1992gm}. We consider this interacting Hamiltonian to impose the boundary conditions on spin 2 and other higher spin fields. We show that the classical evolution of metric and other higher spin fields are governed by the dynamics collective fields and other supplementary fields. Since the dynamics of collective and other supplementary fields has a geometric interpretation in terms of evolution of free fermi droplets in presence of folds \cite{Das:1995gd, Das:2004rx}, our construction therefore provides a phase space description of higher spin gravity in $AdS_3$ in the spirit of LLM \cite{Lin:2004nb}. We also show that for spin 3 gravity one can construct an infinite number of gauge transformations (for a given Hamiltonian) that preserve the asymptotic structure of the bulk spacetime and hence render infinite sequences of conserved charges whose mutual Poisson brackets are zero and thus exhibit an integrable structure.

\paragraph{Discussion :} From the relation between $p_\pm$, $w_{\pm n}$ and folds $f_{\pm i}$ (eqns. (\ref{eq:pfrel}, \ref{eq:pfrel2})) we see that if folds are developed on the upper (for example) fermi surface then all $w_{+n}$s are non-zero in general and are given in terms of $f_{\pm i}$ and vice versa.

For the static solutions of the field equations (\ref{eq:maxeq}) we have
\begin{equation}
    \xi_{\pm} = \text{constant} \quad \text{and} \quad \zeta_{\pm s} = \text{constant} \ \forall \ s.
\end{equation}
In order to maintain the regularity of the Euclidean black hole solution it was shown in \cite{Henneaux:2013dra,Bunster:2014mua,Grumiller:2016kcp, Ojeda:2020bgz} that the chemical potentials $\xi_\pm$ and $\zeta_{\pm s}$ depend linearly on $M-1$ arbitrary integers when $M$ spin fields are turned one. Black holes which are connected to BTZ black holes the above conditions are given by,
\begin{eqnarray}
    \xi_\pm = 2\pi, \quad \zeta_{\pm s} = 0.
\end{eqnarray}
We consider spin 3 black holes whose dynamics is governed by the Hamiltonian (\ref{eq:foldH}). For such black holes these conditions are given by
\begin{align}
    \frac{p_\pm^2}{2} + w_{+1} + V & = 2\pi,
    \label{eq:statsolcond1}\\
    u_{\pm 3}\lb p_\pm + u_{\pm 3}\rb & = 0.
    \label{eq:statsolcond2}
    \end{align}
From the second condition (\ref{eq:statsolcond2}) we see that one trivial solution is $u_{\pm 3} = 0$, i.e. no higher spin modes are turned on. The non-trivial solution is $u_{\pm 3} = -p_\pm$. To find the droplet geometry one has to solve the equations in (\ref{eq:pfrels3}) to find possible values of $f_{\pm i}$. The entropy of the black hole, which is connected to the BTZ black hole, is proportional to $2\pi (p_+ - p_-)$ \cite{Henneaux:2013dra, Bunster:2014mua, Grumiller:2016kcp,  Ojeda:2020bgz}. Therefore, from eqn. (\ref{eq:pfrels3}) we see that the entropy of such black holes is equal to the total area covered by the droplets. However for other black hole solutions (not connected to BTZ) the entropy is no longer equal to the area of the droplets. This is worth mentioning that any arbitrary droplet geometry may not satisfy the regularity conditions at the horizon.

Following the Hamiltonian reduction method developed in \cite{Gonzalez:2018jgp, Grumiller:2019tyl, Ojeda:2020bgz} one can show that the total action is governed by left and right moving chiral bosons for a specific choice of boundary term in (\ref{eq:boundterm}). The  CS action (\ref{eq:CSacn2}) receives contributions only from the boundary degrees of freedom and is given by
\begin{eqnarray}
    I(\phi_{\pm}, \psi_{\pm s}) = \int dt \lB \frac{k}{4\pi} \int d\theta \lb \phi'_\pm \dot{\phi}_\pm + \sum_{s} \psi'_{\pm s} \dot{\psi}_{\pm s} \rb \mp  H^\pm \rB.
\end{eqnarray}
The Euclidean $AdS_3$ partition function for specific boundary conditions, given by the choice of boundary action, can be written as
\begin{eqnarray}
    \cZ = \sum_{\text{classical solutions}} \int [D\Phi] e^{-\beta H}.
\end{eqnarray}
In general this is a difficult problem to address \cite{Maloney:2007ud}. However, we can try to compute the partition function for a given classical solution. In absence of higher spins, one can consider the classical solution to be a BTZ black hole which is given by a constant droplet as discussed in sec. \ref{sec:btz}. Thus excitations about this classical solution correspond to different deformations of the droplet and one has to sum over all such deformations. Expanding $p_\pm$ in Fourier modes we can classify all possible deformations as quantum states in the Hilbert space. The problem was discussed in \cite{Chattopadhyay:2020rle}. It turns out that the partition function is equal to 2D Yang-Mills partition function on torus.  It would be interesting to find the partition function in presence of different higher spin fields in the bulk. In classical theory one needs an infinite number of fields ($w_{\pm n}$) to describe folds on the fermi surface. However, the situation is different in quantum theory. It turns out that $w_{\pm n}$ are not additional degrees of freedom in quantum theory; rather they represent $\cO(1)$ quantum dispersions of the collective fields \cite{Das:1995jw}.

\paragraph{Acknowledgement :}
We are thankful to Nabamita Banerjee, Arghya Chattopadhyay, Sumit Das, Dileep Jatkar, Arnab Rudra, Ashoke Sen for useful discussions. The work of SD is supported by the MATRICS (grant no.  MTR/2019/ 000390, the Department of Science and Technology, Government of India). The work of DM is supported in part by the grant SB/SJF/2019-20/08. Finally, we are indebted to people of India for their unconditional support toward the researches in basic science.

\bibliographystyle{hieeetr}
\bibliography{PlancherelNotes}{}

\end{document}